\input lecproc.sty
\input epsf.sty
\contribution{Variability, Power, and Pair Content of AGN Jets}
\contributionrunning{Variability, Power, and Pair Content of AGN Jets}
\author{Marek Sikora@1, Greg Madejski@2@3, and Mitchell Begelman@4}
\authorrunning{Sikora et al.}
\address{@1Nicolaus Copernicus Astronomical Center, Warsaw
@2NASA/GSFC
@3Dept. of Astronomy, Univ. of Maryland
@4JILA and Department of APS, Univ. of Colorado}
\abstract{
Simultaneous observations of blazars in different spectral regimes
imply that the amplitude of variability depends on the observational
band.  Both in the low energy spectral component and in the high
energy spectral component, the amplitude increases with the photon
energy.  We show that such behavior can be explained in terms of a
two-component model, where the spectra observed during flares are
superpositions of spectra arising from at least two distinct,
spatially separated sources, one which is relatively long-lived
(weeks/months), and another which is short-lived (hours/days).  In the
proposed model, both sources represent dissipative ``events"
(shocks/reconnection sites), propagating along a jet at relativistic
speeds, with the long-lived source ``developing" over much larger
distances in a jet than the short-lived source.  In both of these
sources, the low energy radiation is produced by synchrotron emission,
and the high energy radiation is produced by inverse-Compton (IC).  In
quasars, dominant sources of seed photons for the IC process are
Broad-Emission-Line-Clouds located at distances corresponding to the
propagation distance of the short lived sources and hot dust
distributed over distances corresponding with the extent of the long
lived sources. In BL Lac objects, for both short lived and long lived
sources, the seed photons for the IC process are presumably provided
by local synchrotron radiation.  Using the two-component model for
blazar variability we derive constraints on jet physics in
quasars. Particularly, we demonstrate that if production of X-rays
during rapid flares is dominated by Comptonization of external
radiation, then the upper limits for the total power of jets imply
that the jet plasma is strongly dominated by pairs. A low pair content
would require a low-energy cut-off in the electron energy distribution
and production of X-rays by the SSC process.  The break in the
electron energy distribution should then be imprinted in the soft
$\gamma$-ray band (0.1 - 10 MeV) and, therefore, detailed observations
in this band can provide an exceptional opportunity to study the pair
content in AGN jets.
}
\titlea{1}{Introduction}

Blazars are known to be strongly variable sources in all spectral bands,
from radio, IR, optical, UV, X-rays, up to $\gamma$-rays at GeV and TeV 
energies.  The patterns of variability are usually complex, as they 
cover a wide range of time scales (from hours to years) and depend on 
the spectral band (see, e.g., Ulrich, Maraschi \& Urry 1997).  Most of 
the variability data comes from the low energy bands, collected by 
optical and radio telescopes, often covering a span of tens of years.  
These data show that, on average, the amplitude of variability is 
increasing with photon energy, and the same was recently discovered in 
the high-energy spectral components, both in BL Lac objects and flat
spectrum radio-quasars (Maraschi et al. 1994).  In addition to 
the amplitude-energy dependence within the low energy and the high 
energy spectral components, there is a clear trend observed in quasars 
showing higher variability amplitudes in the high energy spectral component 
than in the low-energy spectral component (see, e.g., Wehrle et al. 1997). 

As we demonstrate in \S 2, these variability properties can be explained in 
terms of a multi-component model, where the blazar spectra observed during 
flares  are superpositions of spectra produced by several spatially 
separated regions of the jet.  In this model, the radiatively active 
regions are represented by dissipative ``events" (shocks/reconnection 
sites), propagating along the jet at relativistic speeds. They develop 
at different distances, and produce radiation which is observed as flares 
with time scales $\sim r/\Gamma^2c $.  The scenario involves synchrotron 
radiation which is responsible for the production of the low-energy spectral 
component and the IC process which is responsible for the high-energy 
spectral component. The seed photons for the latter are  provided by 
local synchrotron radiation as well as  by external sources, and 
the respective processes are denoted  by SSC (synchrotron self-Compton) 
and by ERC (external radiation Compton).  

We show in \S 3 that detailed observations of short term flares in X-ray
and soft $\gamma$-ray bands can be used to determine the relative roles of 
ERC and
SSC radiation mechanisms, and to evaluate the pair content of subparsec 
jets in AGN. 
We summarize the results and predictions of our two-component model in \S 4.

\titlea{2}{The Two-Component Model}

\titleb{2.1}{Characteristic frequencies}

There are several characteristic frequencies where the blazar radiation
components change their spectral slope significantly.  They are related to:
maximum and minimum electron energies; ``cooling'' break  energy (below 
which the time scale of electron energy losses due to radiation processes  
is longer than the time scale of the source propagation); and in the case 
of synchrotron radiation, there is a cut-off due to synchrotron 
self-absorption.  These spectra can also show a high energy cut-off
due to $\gamma$-ray absorption in the pair production process;  
however, for our model parameters, the ambient radiation field is 
transparent even for most energetic $\gamma$-rays.

In the case of synchrotron and SSC components, photons are produced with 
energies scaled by the magnetic field intensity, and since this drops 
with distance, the spectra produced closer to the black hole are expected 
to be shifted to higher frequencies than the spectra produced in the more 
distant regions.  Thus, if the apparent luminosities of flares produced 
closer to the black hole are comparable to or smaller than the apparent 
luminosities of flares produced at larger distances, then the short term 
flares will be observed to have high amplitudes only in the high energy 
tails of the respective spectral components.  This can explain why BL Lac 
objects such as Mkn 421 (Takahashi et al. 1996) show much higher amplitude 
of variability  in X-ray and TeV bands than at lower energies within 
respective spectral components.

In the case of $\gamma$-ray radiation produced by the ERC process, the 
situation is different. The photons are upscattered to energies which 
are proportional to the energy of seed photons, and if the spectrum of the 
external diffuse radiation field doesn't change very much with the 
distance, then for the same electron injection functions, the ERC spectra 
produced at different distances cover the same frequency range.  However, 
there is a quite sharp transition in the spectrum of the diffuse external 
radiation field at a distance $\sim \sqrt {L_{UV}/10^{46} {\rm erg s}^{-1}}$ 
pc, from that dominated by the broad emission lines to that dominated by 
the IR radiation from dust.  Since rapid flares are produced below this 
transition distance, they should have their high energy break at higher energies 
than the ERC spectrum produced at larger distances, approximately by a 
factor $\nu_{UV}/\nu_{IR} \sim 30$.  This can explain why the spectra 
measured by EGRET are steeper during the quiescent phases than those during 
the rapid flares. This is because during the quiescent phases, the radiation 
is produced at larger distances, and thus the high energy break is shifted 
into the EGRET range.

If the electron energy spectrum has a low energy break in the range 
$1 <\gamma_{min}' < 100$, then such a break should be imprinted on the 
ERC(UV) spectrum in the frequency range 1 keV - 10 MeV and in the ERC(IR) 
spectrum in the frequency range up to $\sim 300$ keV.  No spectral break 
related to  $\gamma_{min}'< 100$ can be visible in the synchrotron 
component, because its location is below the frequency at which 
synchrotron-self-absorption becomes important.  Neither is the break 
expected to be imprinted in the SSC component.

A very informative spectral feature is the one related to the electron 
energy, $\gamma_b'$, below which radiative cooling of electrons is 
inefficient.  In the case of electron energy losses dominated by the 
ERC process, this energy is (Sikora et al. 1994)
$$\gamma_b' \simeq {1 \over \Gamma} 
{r \over \xi L_{UV}} {m_e c^3 \over \sigma_T} ,\eqno (1) $$
where $\xi$ is the fraction of the central radiation reprocessed by clouds 
and dust at a distance $\sim r$.  This ``cooling  break'' is imprinted 
in the radiation spectra at energies $\nu_b \propto {\gamma_b'}^2$, i.e. at 
$\nu_{b,EC} \propto r^2$.  However, since the energies of seed photons 
at subparsec distances and at larger distances differ by a factor 
$\nu_{UV}/\nu_{IR} \ge 30$, the location of the break in the spectra of the 
rapid flares does not have to be very different from the location of 
the break in spectra produced at larger distances.  For quasars with 
powerful jets, both are located in the range $1-30$ MeV range.
This explains why in quasars the X-ray spectra are always harder than
$\gamma$-ray spectra, regardless of whether they are detected during rapid 
flares or between them.

\titleb{2.2}{Relative Luminosities}

Since both the ERC luminosity and the synchrotron luminosity scale 
linearly with the injection rate of electrons, the ratio $L_{ERC}/L_S$ 
is predicted to be constant during a flare.  Contrary to this prediction, 
the observations of the best studied blazar, 3C279, show much higher 
amplitudes of variability in the $\gamma$-ray band than in the optical/UV 
band.  Such  variability patterns are predicted by the SSC model, 
where $L_{SSC} \propto L_S^2$.  However, the amplitude ratio observed in 
the January 1996 flare of 3C279 was even larger than that predicted by 
the SSC model (Wehrle et al. 1997).  This can be explained by the 
one-zone ERC model only if the increased injection rate of relativistic 
electrons is accompanied by  increase of the bulk Lorentz factor.
However, at distances of $0.1 -1$ parsec, the jet energetics is 
expected to be dominated by  the kinetic energy flux rather than by 
internal energy of the jet, and any strong dissipative events should 
be followed  by a {\sl deceleration} of the jet, rather than its 
{\sl acceleration}.  

\begfig 0 cm
{\centerline{\epsfysize=6.3truecm\epsfbox{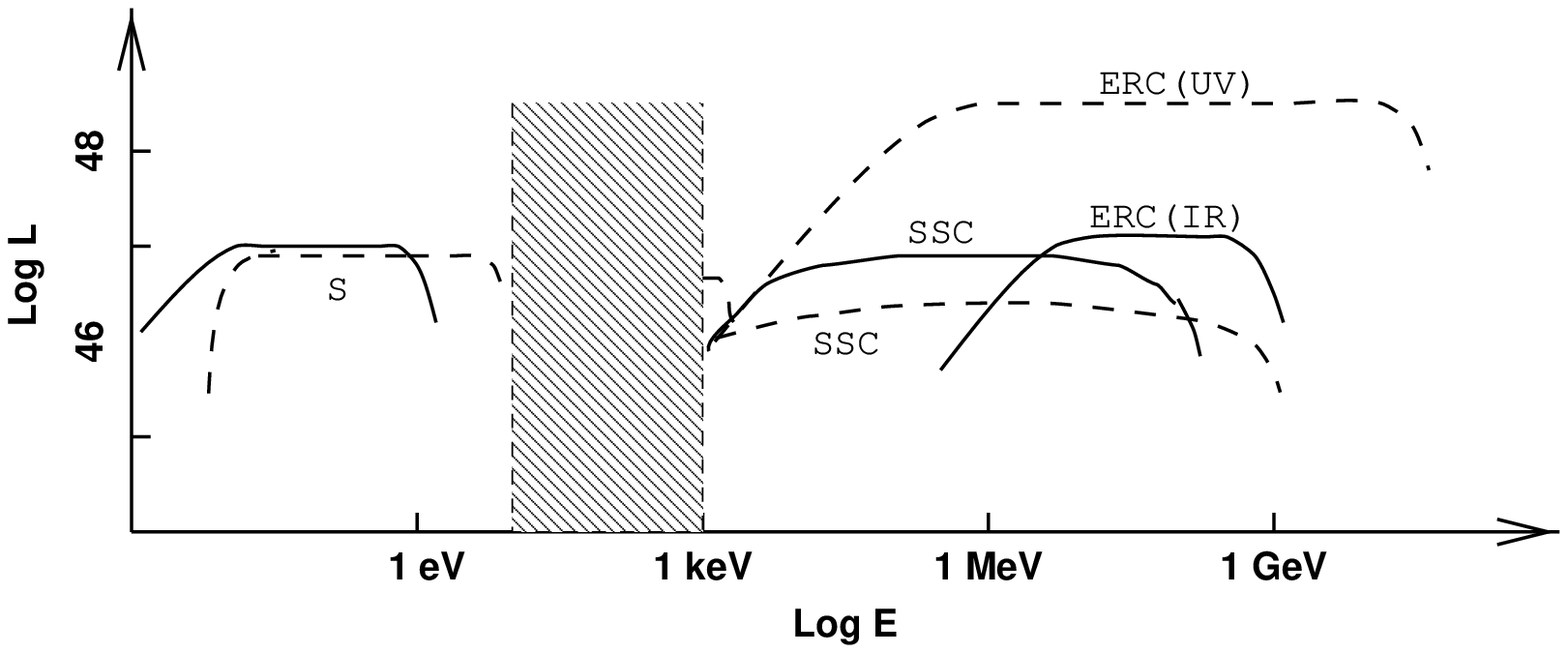}}
\figure{1}{Contribution of different radiation
processes to the spectrum of a blazar.  The solid lines mark the 
radiation produced in a distant zone, while the dashed lines mark 
radiation produced in a near zone.}}
\endfig

The above difficulties can be resolved by using our two-component model, and 
postulating that the more distant dissipative events propagate with a 
lower Lorentz factor than the nearer ones.  The consequences of this are 
schematically illustrated in Fig. 1. There are two spectra: the  one marked by 
a solid line is produced over a distance range $\Delta r_1 = 4.5 - 1.5 = 3$ 
pc by events propagating with the Lorentz factor $\Gamma_1=5$, and the 
one marked by a dashed line is produced over a distance range 
$\Delta r_2 = 0.45 - 0.15 = 0.3$ pc by an event propagating with 
Lorentz factor $\Gamma_2=10$.  The spectra are calculated assuming 
conical geometry of the jet and conservation of magnetic energy flux. The 
opening half-angle of the jet is taken to be $\theta_j \sim 1/\Gamma_2$, 
and the observer is assumed to be located at an angle $\le 1/\Gamma_2$.
The observed variability time scales of such events are 
$t_{fl,1} \simeq \Delta r_1/c \Gamma_1^2 \simeq 160$ days
and $t_{fl,2} \simeq \Delta r_2/c \Gamma_2^2 \simeq 4$ days, 
and the luminosity amplification is respectively 
$\sim \Gamma_2^4$ and $\sim \Gamma_1^4$.  In our example, the 
energy dissipation rate in both events is the same, and is calculated 
assuming that the apparent luminosity of the rapid flare is 
$\sim 10^{48} {\rm erg s}^{-1}$.  Electrons are assumed to be injected 
with the same distribution in both events, with the power index 
$q = 2 \,(Q \propto {\gamma'}^{-q})$ and maximum energy $\gamma_{max}'=10^4$. 
The presented spectra are calculated for 
$\xi L_{UV} = L_B = 10^{45} {\rm erg s}^{-1}$, where 
$L_B \simeq ({B'}^2/8\pi)c\Gamma^2 \pi a^2$ is the flux of magnetic energy
carried by the jet plasma, and $a = \theta_j r$ is the source size.

One can see from Figure 1 that during the rapid flare, a comparable 
contributions to the X-ray band can come from  ERC and SSC radiation produced 
in this flare and from the SSC radiation produced at larger distances.
Hence, the latter dilutes the X-ray part of the rapid flare and this 
explains why the fast flares show lower fractional variability amplitude 
in the X-ray band than in the $\gamma$-ray band.  Our example also shows 
that in the two-component scenario, the amplitude of the $\gamma$-ray 
variability can be much larger than the amplitude of the optical flare.
This is a consequence of the fact that ratio of the ERC luminosity to the    
synchrotron luminosity depends strongly on $\Gamma$.  Then, for the 
rapid flare produced closer to the black hole and  moving faster, this 
ratio is much higher than for radiation produced at larger distances.

We have also marked in Figure 1 the bump at $\sim$ 1 keV.  It is predicted to
result from Comptonization of UV radiation by cold electrons in a jet 
(Begelman \& Sikora 1987; Sikora et al. 1997) and is not confirmed by observations.  Production of 
such a bump can be avoided near the black hole by assuming that the jet is 
not  
yet fully developed at this point.  But at the distances where the strong 
nonthermal flares are produced, the jet must be already fully developed 
(collimated, accelerated, and mass loaded), and any presence of cold 
electrons should be imprinted in the excess of radiation above the 
power-law X-ray spectra in the soft X-ray band, provided that the 
production of X-rays down to lowest energies is dominated by Comptonization 
of UV radiation.  Such a bump can be partially 
hidden by the SSC component produced at larger distances. Also, there can 
be a significant contribution,
not marked in the Figure, from Comptonization of the dust radiation 
by electrons in the ``near'' event.
Finally, the lack of the ``soft X-ray bump'' can 
result from the cutoff of electron distribution at $\gamma' \gg 1$, but 
then the rapid X-ray flares correlated with the $\gamma$-ray flares 
must  be produced by the SSC process.  However, flares produced by the 
SSC process are predicted to have much softer spectra than observed.

\titlea{3}{Power and Pair Content}

The spectral shapes of $\gamma$-ray flares reflect roughly the power-law 
injection function of relativistic electrons, with a power-index
$q \sim 2$.  If the electrons are injected with the same slopes down to 
the lowest energies, then the average energy of an injected electron is   
$\bar \gamma'$ is $\sim 10$.  The electrons injected  at energies 
$> \gamma_b'$ are cooled very efficiently, and produce $\gamma$-ray 
spectra with the energy index $\alpha_{\gamma} =q/2 \sim 1$, while the 
electrons injected at lower energies reradiate only a fraction $(r/c)/t_e$ 
of their energy, and thus produce spectra with the energy index 
$\alpha_X \sim 0.5$, where $t_e$ is the time scale of the electron 
radiative energy losses.  The rate at which the jet energy must be 
converted to relativistic electrons producing apparent $\gamma$-ray luminosity 
$L_{\gamma}$ is
$$L_e \simeq L_e' \simeq  L_{\gamma}'/\eta_{\gamma} \simeq 
L_{\gamma}/\eta_{gamma}\Gamma^4 , \eqno (2)$$
where $\eta_{\gamma}$ is the fraction of injected energy reradiated by 
electrons in the $\gamma$-ray band.  (Note that $L_e = L_e'$ because 
for isotropic electron injection the angle averaged momentum in the source frame is zero and electron energies are Lorentz-transformed 
like  time intervals.)  

For a given ratio of electron to proton densities, $n_e/n_p$, 
the minimal energy requirement for the energy flux of the jet to 
provide $L_e$ can be obtained by assuming that electrons are injected 
at the expense of the bulk-kinetic energy flux, $L_K$, dominated by 
cold protons. If the bulk velocity is $\beta_0$ and the shock, where 
the jet kinetic energy is dissipated, moves at speed $\beta < \beta_0$ and 
both are relativistic, then energy flux of the matter moving through the 
shock front is
$$L_{in} = L_K (\beta_0 - \beta) \simeq 
{L_K \over \Gamma^2} {\Delta \Gamma \over  \Gamma} \eqno (3) $$
This together with eq. (2) gives
$${\Delta \Gamma \over \Gamma} =  {1 \over \eta_e \eta_{\gamma} \Gamma^2}
{L_{\gamma} \over L_K}, \eqno (4)$$
where  $\eta_e \equiv L_e/L_{in}$ 
Noting that $n_e m_e \bar \gamma' \Gamma = \eta_e n_p m_p \Delta \Gamma$
and using  equation (4) we obtain
$$ \bar \gamma' = \eta_e {\Delta \Gamma \over \Gamma} {n_p m_p\over n_e m_e} = 
{1 \over \eta_{\gamma}} {n_p m_p\over n_e m_e} {L_{\gamma} \over \Gamma^2 L_K}
\simeq 600 {n_p/n_e  \over (\eta_{\gamma}/0.3) (\Gamma/10)^2} 
{ L_{\gamma,48} \over L_{K,47}} \eqno (5) $$   
This can be reconciled with low values of $\bar \gamma'$ ($\sim 10$ for 
$q \simeq 2$) obtained
under the assumption that there is no low energy break in the electron 
injection function only if the jet plasma is strongly dominated by pairs.  
Otherwise the jet kinetic luminosities must be extraordinary high or 
the electron injection function must have a low-energy cutoff at 
$\gamma_{min}' \sim 100 (n_p/n_e)$. It is worth noting that the same conclusion
has been reached from analyses of energetics of VLBI sources 
(Ghisellini et al. 1992; Celotti \& Fabian 1993; Reynolds et al. 1996). 
The stage of the low-energy electron distribution 
can be verified directly by observations in the X-ray band and in the soft 
$\gamma$-ray band ($0.1 - 10$ MeV), where the  low-energy break should be 
imprinted in the ERC spectral component.

We would like to emphasize that the constraints on e-p jets obtained from
equation (5) can be somewhat eased by allowing that $L_K$ is strongly 
modulated by the central source.  In particular, it is possible that on 
subparsec scales the jet is not continuous, and its structure is marked 
by a sequence of ``clouds'' ejected from time to time from the central 
source.  Then the $L_K$ used in eq. (5) represents the flux of energy 
carried by a single cloud which can be much larger than the time-averaged 
kinetic energy flux deduced from energetics of large scale radio structures 
or from energetics of the central engine. 

\titlea{4} {Conclusions}

To explain the variability patterns of blazars, we suggest a two-component 
model of a jet, where the spectra observed during flares are
superpositions of spectra arising from at least two distinct,
spatially separated sources, one which is relatively long-lived
(weeks/months), and another which is short-lived (hours/days). The 
predictions of the two-component  model are that:

\noindent
--- Rapid flares (days in flat spectrum radio-quasars; hours in BL Lac objects)
 should 
be fully observed (only marginally diluted) in high energy tails of both 
the synchrotron  and the Compton spectral components, and these two
components should be correlated.  
These flares are expected to be significantly diluted in 
the low energy parts of the synchrotron and Compton spectral components 
by radiation from the distant regions; 

\noindent
--- Long flares  (= high states lasting for weeks/months) 
have negligible contribution to the high energy tails of both synchrotron and
Compton components. They, however, can be visible as a background
quasi-steady radiation, on which sporadic rapid flares are superimposed.  
In the spectral bands where high energy tails of distant sources 
contribute, the flares cause hardening of the observed spectra,
simply because the spectra of rapid flares are shifted to higher 
frequencies in comparison with spectra of distant sources;

\noindent
--- The frequency shift between spectra of the distant and the near regions 
is, in general, different for the synchrotron and Compton components.  
This means that for the ERC models, the synchrotron  and Compton peaks 
are in general dominated by the radiation produced by regions spatially 
separated in a jet, and, therefore, the assumption about co-spatiality 
of plasmas producing both peaks, often used to derive physical parameters 
in a jet (Ghisellini, Maraschi, \& Dondi 1996; Sikora et al. 1997), can be 
overly simplistic;

\noindent
--- The higher amplitudes of $\gamma$-ray flares compared to optical/UV
flares can be explained by assuming that the distant dissipative events
propagate more slowly than the closer ones.  This is because the ratio
of the Compton luminosity to the synchrotron luminosity in the the ERC model 
depends very strongly on $\Gamma$;

\noindent 
--- Production of X-rays during the intervals {\sl between} flares can 
be dominated by the SSC process, while X-ray flaring by a factor $>2$ 
is most likely due to the ERC process with the seed photons provided 
by broad emission line clouds and dust;

\noindent
--- The observed spectra of high energy flares in quasars indicate that the 
jets are dominated by pairs, i.e. $n_e/n_p \gg 1$.
The electron-proton jet models require that the electron injection function 
has a low-energy cutoff at $\gamma_{min}' \sim 100$. This should be
imprinted in the spectra of ERC radiation components at  energies 
$\sim 10$ MeV, while X-rays should be produced entirely by the SSC 
mechanism. Our model then predicts steeper X-ray spectra than are observed.

\smallskip
MS acknowledges support from KBN grant 2P03D01209 and MCB acknowledges
support from NSF grant AST-9529175 and NASA grant NAG5-2026.

\begrefchapter{References}
\ref Begelman, M.C. \& Sikora, M. 1987, ApJ,322, 650
\ref Celotti, A. \& Fabian, A.,C. 1993, MNRAS, 264, 228 
\ref Ghisellini, G., Celotti, A., George, I.M., \& Fabian, A.C. 1992,
MNRAS, 258, 776
\ref Ghisellini, G., Maraschi, L., \& Dondi, L. 1996, A\&AS, 120, C503
\ref Maraschi, L. et al. 1994, ApJ, 435, L91
\ref Reynolds, C.S., Fabian, A.C., Celotti, A., \& Rees, M.J. 1996, MNRAS,
283, 873
\ref Sikora, M., Begelman, M. C., \& Rees, M. 1994, ApJ, 421, 153
\ref Sikora, M., Madejski, G. M., Moderski, R., \& Poutanen, J. 1997, ApJ, 
484, 108
\ref Takahashi, T., et al. 1996, ApJ, 470, L89
\ref Ulrich, M.-H., Maraschi, L. \& Urry, C. M., 1997, Ann. Rev. Astron. 
Astrophys., in press
\ref Wehrle, A., et al. 1997, submitted to ApJ

\endref

\byebye